\documentclass[conference,compsoc,final]{style/IEEEtran}
\ifCLASSOPTIONcompsoc
  \usepackage[nocompress]{cite}
\else
  \usepackage{cite}
\fi

\usepackage[pdftex]{graphicx}
\graphicspath{{./figures/}}
\DeclareGraphicsExtensions{.pdf,.jpeg,.png}

\ifCLASSINFOpdf
\else
\fi
\usepackage{amsmath}
\usepackage{algorithmicx}
\usepackage{algpseudocode}
\usepackage{algorithm}

\usepackage{multirow}
\usepackage{balance}  
\usepackage{graphics} 
\usepackage{times}    
\usepackage{balance}  
\usepackage{graphicx}
\usepackage{subfig}
\usepackage{epstopdf}
\usepackage{rotating}
\usepackage{xspace}
\usepackage{enumitem}

\newcommand{\myalgo}{$PTR$\xspace}

\newcommand{\gardenhose}{\ensuremath{{\cal G}}\xspace}
\newcommand{\tweets}{\ensuremath{{\cal T}}\xspace}
\newcommand{\tweetspos}{\ensuremath{{\cal T}_{c+}}\xspace}
\newcommand{\tweetsneg}{\ensuremath{{\cal T}_{c-}}\xspace}
\newcommand{\tweetsmenloc}{\ensuremath{{\cal T}_{ML}}\xspace}
\newcommand{\tweetsuserloc}{\ensuremath{{\cal T}_{UL}}\xspace}
\newcommand{\users}{\ensuremath{{\cal U}}\xspace}
\newcommand{\userspos}{\ensuremath{{\cal U}_{c+}}\xspace}
\newcommand{\usersneg}{\ensuremath{{\cal U}_{c-}}\xspace}
\newcommand{\userloc}{\ensuremath{{\cal U}_L}\xspace}
\newcommand{\topics}{\ensuremath{{\cal C}}\xspace}
\newcommand{\htags}{\ensuremath{{\cal H}}\xspace}

\usepackage[hide]{style/chato-notes}
\usepackage{hyperref}

\begin{document}
%

\title{Sentiment-enhanced Multidimensional Analysis of Online Social Networks:\\
Perception of the Mediterranean Refugees Crisis\\
{\small [UNDER REVIEW]}}

\author{
\IEEEauthorblockN{Mauro Coletto\IEEEauthorrefmark{1}\IEEEauthorrefmark{2}, Claudio Lucchese\IEEEauthorrefmark{2}, Cristina Ioana Muntean\IEEEauthorrefmark{2},\\Franco Maria Nardini\IEEEauthorrefmark{2}, Andrea Esuli\IEEEauthorrefmark{2}, Chiara Renso\IEEEauthorrefmark{2}, Raffaele Perego\IEEEauthorrefmark{2}}
\IEEEauthorblockA{\IEEEauthorrefmark{1} IMT School for Advanced Studies Lucca}  
\IEEEauthorblockA{\IEEEauthorrefmark{2} ISTI - CNR Pisa\\
email: \{name.surname\}@isti.cnr.it}}


%


\maketitle

\begin{abstract}
We propose an analytical framework able to investigate discussions about polarized topics in online social networks from many different angles.
The framework supports the analysis of social networks along several dimensions: time, space and sentiment.
We show that the proposed analytical framework and the methodology can be used to mine knowledge about the perception of complex social phenomena. 
%
We selected the \textit{refugee crisis} discussions over Twitter as the case study.  
This difficult and controversial topic is an increasingly important issue for the EU.
The raw stream of tweets is enriched with space information (user and mentioned locations),
and sentiment (positive vs.\ negative) w.r.t.\ refugees.
Our study shows differences in positive and negative sentiment in EU countries, in particular in UK, and by matching events, locations and perception it underlines opinion dynamics and common prejudices regarding the refugees.

\end{abstract}


%
\IEEEpeerreviewmaketitle

\section{Introduction}
\label{sec:introduction}



\begin{figure}[t]
	\centering
		\includegraphics[width=1\columnwidth]{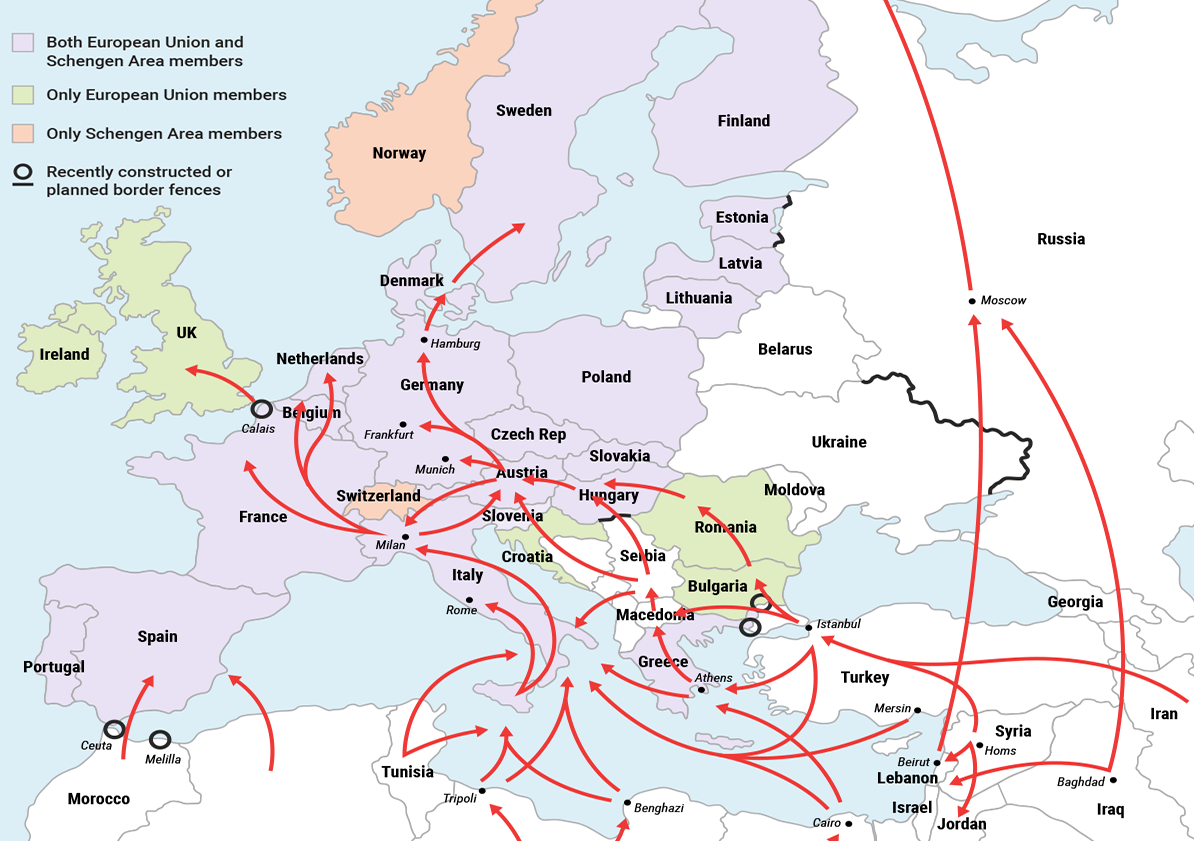}
	\caption{The routes to European countries (source Business Insider - Europol, Reuters, Washington Post, AFP, ICMPD).}
	\label{fig:route}
\end{figure}

We are recently witnessing one of the largest movement of migrants and refugees from Asian, African and Middle-east countries towards Europe. The United Nations High Commissioner for Refugees (UNHCR) estimates
one million of refugees arrived to the Mediterranean coasts in 2015 mainly from Syria (49\%), Afghanistan (21\%) and Iraq (8\%). Figure~\ref{fig:route} reports main routes to EU coasts and to northern Europe. 
%
The largest wave of arrivals started in August 2015 following a main route through Turkey, Greece, Macedonia, Hungary and Austria to Germany, France, UK and other northern European countries. Since then, this phenomenon has been in spotlight of the media, which have reported an increasing number of events related to migrants, such as the additional border controls established by Hungary, Austria and Germany, the several incidents happened to refugees, or the touching photo of the young Syrian boy found dead on the seashore in Turkey in September 2015. 
  
The implications of this refugee crisis are several and complex. 
The whole phenomenon is nowadays object of a heated and polarized debate. 
Understanding how the debate is framed between governmental organizations, media and citizens may help to better handle this emergency.
Through the analysis of the Twitter online social network, we address the following questions:
``How is the European population perceiving this phenomenon? What is the general opinion of each country? How is perception influenced by events? What is the impact on public opinion of news related to refugees? How does perception evolve in time in different European countries?"

Social media may help in answering these relevant questions but the volume of messages exchanged is massive and the extraction of sentiment is challenging.  
Basic analyses of the phenomenon through Twitter have been already performed by media in a simplistic way, mainly through manual analysis of content, news and hashtags for small samples of tweets. For instance, the usage of terms \textit{refugees} and \textit{migrants} have been compared\footnote{https://storify.com/ImagineEurope/what-is-associated-with-europe} in offline and online media. The former indicates someone forced to leave her country to avoid war or imprisonment,  the latter is instead someone moving from his/her country searching for better living conditions. Additional basic quantitative analyses on Twitter hashtags about refugees have been performed, i.e. the \textit{\#welcome refugees} and \textit{\#germany} hashtags are used mainly from outside Germany\footnote{http://orientalreview.org/2015/09/21/who-is-twitter-luring-refugees-to-germany/}.
These are a few examples on simple analyses performed. However a more comprehensive work trying to analyze the perception of users about these events, shaped across places and time, is still missing. 

In this work we propose an analytical framework to interpret social trends from large tweet collections by extracting and crossing information about the following three dimensions: time, location and sentiment. We describe the methodology to: 1) extract relevant spatial information, 2) enrich data with the sentiment of the message and of the user (retrieved in an automatic iterative way through Machine Learning), 3) perform multidimensional analyses considering content and locations in time. The approach is general and can be easily adapted to any topic of interest involving multiple dimensions. It is scalable due to the automatic enrichment procedure. For the scope of this paper we use our framework to outline the European perception of the refugee crisis.

\section{Related Work}
\label{sec:related}

There is a significant increase of interest in collecting and analysing geo-located data from online social networks (OSNs). Several works study different aspects of the geographical dimension of OSNs, a broad study on this argument is reported in \cite{SMML10}. The authors propose a framework to compare social networks based on two new measures: one captures the geographical closeness of a node with its network neighborhood and a clustering coefficient weighted on the geographical distance between nodes. 
Twitter geo-located posts are studied in \cite{TGW12} to understand how Twitter social ties are affected by distance. Linked users are identified as ``egos'' and ``alters'' and the distance between them is analyzed by considering the correlation with the air travel connection distance and with national borders and languages. 
An analogous objective is the focus of \cite{KKNG12} where the authors infer the location of 12 million Twitter users in a world-wide dataset. Differently from the previous paper, they study the correlation between the Twitter population and the socio-economic status of a country, suggesting that high developed countries are characterized by a larger Twitter usage.  
The geographical properties of Twitter are also useful to study the movements of people and migration phenomena.
A study of mobility  using geo-located Twitter messages is presented in \cite{HSBSKR14}. The authors introduce a detailed study aimed at estimating international travelers based on the country of residence. They identify a number of characteristics including radius of gyration and mobility rate to describe the traveling phenomena thorough the Twitter lens. 
Authors in \cite{ZGVWS14} show how the analysis of 500,000 geo-located Twitter users may help to predict the migration turning points and to better understand 
migration in OECD countries. The authors estimate the migration rate of users moving from one ``home'' country  to another country. The reported results depict some interesting trends such as the decrease of migration from Mexico to US, consistent with official estimations. 

Twitter is also exploited to better understand how the communication flows during political movements and events~\cite{conover2013geospatial}. This work studies  Twitter data covering the birth and maturation of the American anti-capitalist movement \emph{Occupy Wall Street}. The authors analyze the geo-spatial dimension of tweets in combination with the communication dimension building a geographic profile for the communication activity of the movement. 
An extensive analysis of these data produced many interesting results. For example, it appears that proximity to events plays a major role in determining which content receives the most attention in contrast to the stream of domestic political communication. 


Using Twitter for opinion mining and user polarization is a vast subject \cite{pang2008opinion}. The sentiment analysis methods proposed are many, mainly based on dictionaries and on learning techniques through unsupervised \cite{pak2010twitter} and supervised methods (lexicon-based method \cite{taboada2011lexicon}) and combinations \cite{kolchyna2015twitter}.
Opinion mining techniques are widely used in particular in the political context \cite{adamic2005political} and in particular on Twitter \cite{coletto2015electoral}.
Recently new approaches based on polarization, controversy and topic tracking in time have been proposed \cite{Garimella, coletto2016}. The idea of these approaches is to polarize users of a social network in groups based on their opinion on a particular topic and tracking their behavior over time. These approaches are based on network measures and clustering \cite{Garimella} or hashtag classification through probabilistic models \cite{coletto2016} with no use of dictionary-based techniques. 

The novelty of our proposal compared to the state-of-the-art approaches is mainly the fact that 
we propose an analytical framework to study a mass event from Twitter messages as a combination of three dimensions: time, space and sentiment. The sentiment analysis method adopted is efficient in track polarization over Twitter
 w.r.t.\ other more generic methods.
Differently from many approaches studying migration, we do not base our analyses on the change of location of Twitter users to measure the flow of individuals through space, but rather we aim at understanding the impact on the EU citizens perception  of migrants movements.

%



\section{The Analytical Framework}
\label{sec:methodology}

In this section we introduce the analytical framework detailing the data collection phase and the analytical dimensions, namely the spatial, temporal and sentiment dimension.  The final multi-dimensional dataset can be analyzed and queried along these axes and, more interestingly, on combinations between them. The main datasets and the notation used is summarized in Table \ref{tab:notation}.

\begin{table}
\footnotesize
\caption{Notation summary.}\label{tab:notation}
\begin{tabular}{ c | p{4.7cm} || r }
 {\bf Symbol} & {\bf Description} & {\bf \# Total} \\
  \hline
  \gardenhose & Collected English tweets & 97,693,321 \\
  \tweets & Tweets related to the refugee crisis & 1,238,921 \\
  \tweetspos & Positive sentiment tweets & 459,544 \\
  \tweetsneg & Negative sentiment tweets & 387,374 \\
  \tweetsmenloc & Tweets with mentioned location & 421,512 \\
  \tweetsuserloc & Tweets with user location & 101,765 \\
  \users & Users & 480,660 \\
  \userspos & Users with positive sentiment & 213,920 \\
  \usersneg & Users with negative sentiment & 104,126 \\
  \userloc & Users with country location & 47,824 \\
\end{tabular}
\end{table}



We used the Twitter streaming API to collect English tweets data under the \textit{Gardenhose} agreement (10\% of all tweets in Twitter) in period from mid August to mid Sept 2015, noted with \gardenhose, out of which we selected the tweets related to the refugee crisis topic, called the {\em relevant} tweets (denoted as \tweets). We did this by manually choosing a subset of 200 hashtags frequently used and specifically related to refugees in the period of analysis. 
From \tweets we extracted information about three main dimensions: \textbf{ spatial, temporal, and sentiment-based}, resulting in the set of users and tweets as reported in Table~\ref{tab:notation}. 


\subsection{Spatial and Temporal Dimensions}

For each tweet we extract two kinds of spatial information if present: the \textit{user location} of the person posting the message and the \textit{mentioned locations} within the tweet text. 
The \textit{user location} is structured in two levels, the city (if present) and the country. The user city is identified from the \textit{GPS coordinates} or \textit{place} field when available. Since GPS and {\em place} data are quite rare (about 3.5K in all dataset) we used the free-text \textit{user location} field to enrich location metadata. 
We identified locations in the user generated field
based on location data from the Geonames\footnote{http://www.geonames.org/} dictionary which fed a parsing and matching heuristic procedure.
This technique provides high-resolution, high-quality geo-location in presence of meaningful user location data \cite{onnela2011geographic}. The user country  is collected in a similar way and when not explicitly present we infer from the city.
The \textit{mentioned locations} in the text are also represented at city and country level, and they are extracted from tweets' text with the same heuristic procedure as for user location.


We limit our analysis to the perception and sentiment of European citizens.
For the mentioned locations we are also interested to the countries involved in the migration crisis.
The numbers of tweets with user location \tweetsuserloc and mentioned locations \tweetsmenloc are reported in Table \ref{tab:notation}. 

Finally, we extracted the publishing time of each tweet and the period of time when each user was active. This information is necessary to study the evolution of the migrant crisis phenomenon
and users' perception over time.

\subsection{Sentiment Dimension}

We are interested in understanding if the user has a positive feeling in welcoming the migrants or if he/she mainly expresses negative feelings (fear, worry, hate).
Therefore, the dataset is enriched with information about the sentiment for both of tweets and users.

We consider two polarized classes $c \in \topics$: \textit{pro refugees} ($c_+$) and \textit{against refugees} ($c_-$). 
We implemented and modified the algorithm $PTR$~\cite{coletto2016} to assign a class to each polarized tweet and to each polarized user in an iterative way by considering his/her tweets and the hashtags contained.
The approach proposed in \cite{coletto2016} is suitable to track polarized users according to a specific topic which is in our case the ``refugees phenomenon''. 

\begin{algorithm}[t!]
\small
\caption{\label{alg:iterative} \myalgo Algorithm}
\begin{algorithmic}[1]
\Require a set of users \users, their tweets \tweets with hashtags \htags,
\Statex \hspace{2em} a set of hashtags $H_c^{0}$ for each class $c \in \topics$
\Ensure Classification of users $U_c$ and hashtags $H_c$
\Procedure{\myalgo}{\; $\{H_c^{0}\}_{c \in \topics}$\; }
\State $\tau \gets 0$
\Repeat
\Statex \Comment{Classify tweets on the basis of the used hashtags}
\State $\left\{T_{c}^{\tau} \right\}_{c \in \topics} \gets \mbox{\sc TwClass}\left(  \left\{ H_{c}^{\tau} \right\}_{c \in \topics}, \cal{T} \right)$
\Statex \Comment{Classify users on the basis of the written tweets}
\State $\left\{U_{c}^{\tau} \right\}_{c \in \topics} \gets \mbox{\sc UsClass}\left(\left\{T_{c}^{\tau}\right\}_{c \in \topics}, \cal{U}   \right)$
\Statex \Comment{Find better hashtags on the basis of $U_c^{\tau}$}
\State $\{H_c^{\tau+1}\}_{c \in \topics} \gets \mbox{\sc HtClass}(\{U_c^{\tau}\}_{c \in \topics})$
\State $\tau \gets \tau+1$
\Until{ {\em convergence} }
\State \textbf{return} $\{{\cal{U}}_c^{\tau}\}_{c \in \topics}, \{{\cal{T}}_c^{\tau}\}_{c \in \topics}, \{H_c^{\tau}\}_{c \in \topics}$
\EndProcedure
\end{algorithmic}
\end{algorithm}

The Alg.~\ref{alg:iterative} illustrates the procedure. As  input the algorithm receives an initial set of polarized hastags $\left\{ H_{c}^{0} \right\}$ (initial \textit{seed}) for each class $c$ and the collection of relevant tweets \tweets.
The initial seeds have been selected by analyzing the most frequent among about 95 K unique hashtags:

{\small $H_{c_+}^{0}=$} {\footnotesize \texttt{\#refugeeswelcome \#refugeesnotmigrants \#welcomerefugees}}

{\small $H_{c_-}^{0}=$} {\footnotesize \texttt{\#refugeesnotwelcome \#migrantsnotwelcome \#norefugees}}

The initial seed $H_{c_+}^{0}$ is used in 36K tweets, whereas hashtags in class $c_-$ are used in only 2K tweets. One of the benefits of $PTR$ is that after only a few iterations the results are less dependent on the size of the original seed, correcting the unbalanced number of occurrences of the seed by class $c$ in the data.

The internal functions of Alg.~\ref{alg:iterative} are defined as follows:

\begin{itemize}[leftmargin=*]
\item $\mbox{\sc TwClass}$: a tweet is polarized to one class $c$ only if it contains only hashtags of one
class $\left\{ H_{c} \right\}_{c \in \topics}$.

\item $\mbox{\sc UsClass}$: a user is polarized to one class $c$ only if 
his polarized tweets of class $c$ are at least twice the number of his polarized
tweets of any other class.

\item $\mbox{\sc HtClass}$: an hashtag $h$ is assigned to one class $c$ if $S_c(h)> \beta \cdot S_{c'}(h) \quad \forall c'\neq c$, where
$S_c(h)$ is the conjunct probability of observing $h$ in tweets polarized to class $c$, and 
not observing $h$ in the tweets polarized to other classes.
(For our experiments $\beta=0.005$ was chosen after empirical evaluation.)

\end{itemize}


The procedure adds information about polarization of the users by polarized hashtags extension through the analysis of all the tweets written by an already polarized users and not only the polarized tweets.
The iterative procedure has been run 4 times until convergence has been reached. We excluded from the hashtags retrieved by $PTR$ all the hashtags which directly mention a city or a country. This is to keep the sentiment value independent by the location in computing of the polarization. The combination of location and sentiment is done by crossing the space and sentiment. 

Alg.~\ref{alg:iterative} extracts new hashtags at each iteration. We report the most relevant retrieved hashtags in addition to seed ones after the final iteration of the algorithm for
each class $c$:

{\small $H_{c_+}^{\tau=final}$} : {{\footnotesize \texttt{\#campliberty
\#health
\#humanrights
\#marchofhope
\#migrantmarch
\#refugee
\#refugeecrisis
\#refugeemarch
\#refugeescrisis
\#sharehumanity
\#solidarity
\#syriacrisis
\#trainofhope}}

{\small $H_{c_-}^{\tau=final}$} : {\footnotesize \texttt{
\#alqaeda
\#guns
\#illegalimmigration
\#illegals
\#invasion
\#isis
\#islamicstate
\#justice
\#migrant
\#migrantcrisis
\#muslimcrimes
\#muslims
\#no2eu
\#noamnesty
\#nomoremigrants
\#nomorerefugees
\#patriot
\#quran
\#stoptheeu
\#taliban
\#terrorism}}

From the analysis of the extracted hashtags we can see that people with a positive sentiment prefer to use the term \textit{refugees}, while people with a negative sentiment refer to them as \textit{migrants}, thus minimizing the fact that they are escaping war and persecution.  Users with a negative sentiment frequently use \textit{refugees} and the Islamic religion together, somehow correlating, in a prejudicial way, refugees with Islam and terrorism. Finally, we observe that individuals with negative sentiment are often patriotic and not pro Europe.

%
%
%
Note that the algorithm, may classify both users and tweets as non polarized, thus
favoring accuracy of truly polarized content.
The polarization algorithm was able to assign the sentiment to 68\% of the tweets and to 66\% of the users in our dataset. 
Regarding EU-geolocated tweets and users,
the algorithm assigned the sentiment to 73\% of tweets and to 71\% of the users.

%
%



\section{Refugees Crisis Perception Analysis}
\label{sec:experiments}
%

Our study is driven by the analytical questions below: 

\textbf{AQ1}: Is it possible to capture the dynamic of the refugees migration?

\textbf{AQ2}: What is the sentiment of users across Europe in relation to the refugee crisis? What is the evolution of the perception in countries affected by the phenomenon?

\textbf{AQ3}: Are users more polarized in countries most impacted by the migration flow?


\subsection{Spatial and temporal analysis}

We navigate our multidimensional dataset by first analysing the spatial and temporal dimensions to answer {AQ1}. These analysis can quantify the volumes of relevant  Twitter messages based on the countries of the users and the country mentions, since these volumes are strong indicators of real-world  events \cite{weng2011event}. 
Figure~\ref{fig:tweets-country} depicts the total number of tweets for the 20 most active countries.
 
\begin{figure}[h]
	\centering
		\includegraphics[width=0.95\columnwidth]{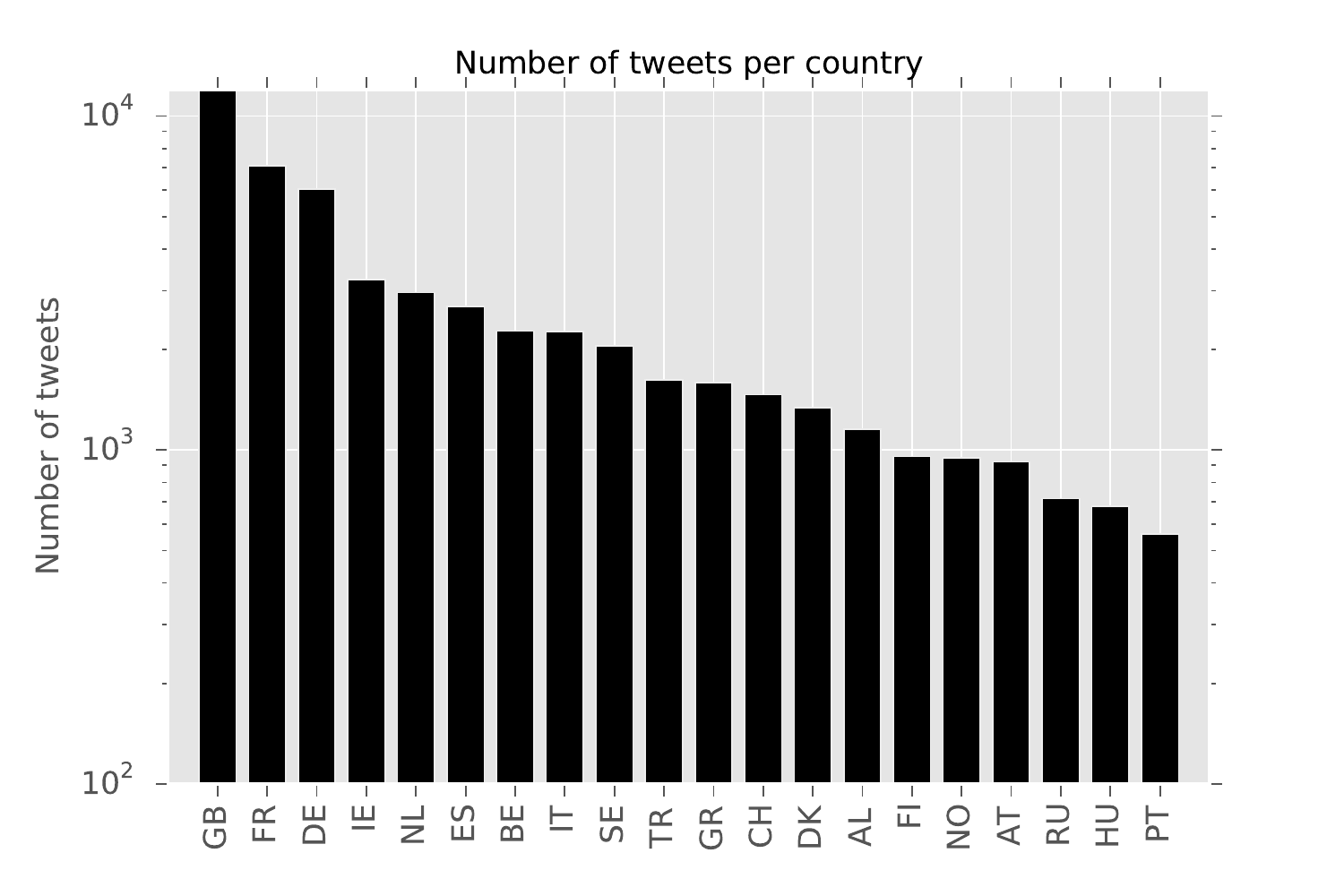}
	\caption{\tweetsuserloc per top-20 countries in log scale.}
	\label{fig:tweets-country}
\end{figure}

\begin{figure}[h]
	\centering
		\includegraphics[width=0.95\columnwidth]{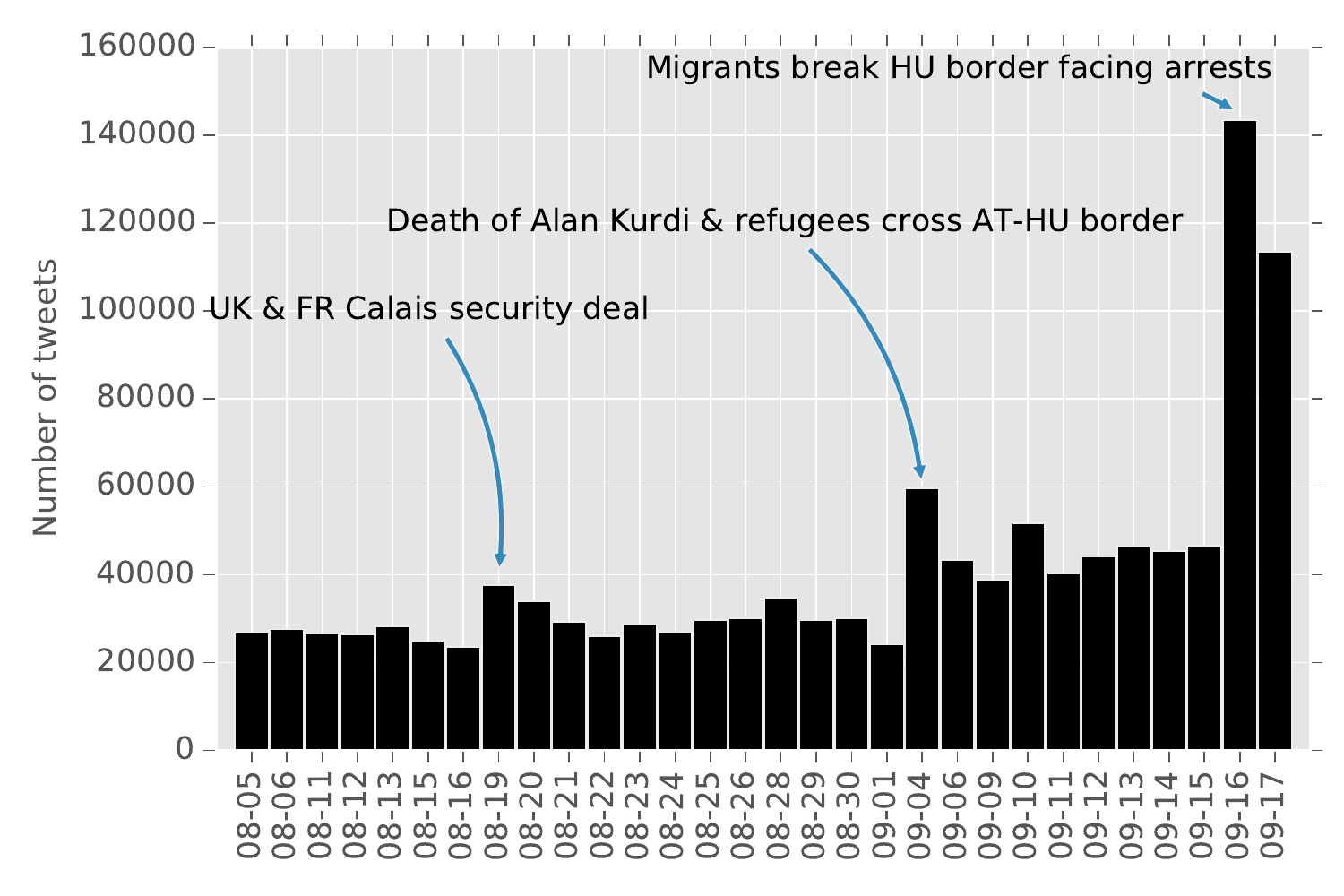}
	\caption{\tweets per day and top pieces of news.}
	\label{fig:tweets-day}
\end{figure}


Since the dataset is in English most of the tweets 
($56.1$\%) come from users located in United Kingdom (UK), therefore in Section~\ref{sec:uk} we focus our analysis on UK. Nevertheless, a significant fraction of the data
come from other countries, e.g., France (FR) accounts for $6.9$\% of the tweets and Germany (DE) 
accounts for $5.9$\%. Without loss of generality, our methodology can be extended to other languages
by simply extending the seed hashtags used in the sentiment dimension construction. 
As far as the mention location is concerned, we see that users from 51 countries mention 154 countries Europe, Asia and Africa.
This is analyzed in further detail in Section~\ref{sec:uk}.





Figure~\ref{fig:tweets-day} illustrates tweets volumes along the temporal dimension, therefore relating volumes to events, summarized in Table \ref{tab:events}. 
This clearly includes all the tweets in \tweets and not only the geo-located ones. We observed significant volume peaks in
days August 19, September 4 and September 19. As we can see from the table, these days match the major events since the UK and France security deal signed on the 18 of August regarding Calais, the drowned Syrian boy found in a beach, Hungary takes refugees to Austrian border by bus in days 2 to 4 September,  
and migrants breaking through Hungarian border on September 16.

\begin{table}
\scriptsize
\caption{Major events during the observation period as reported by UK Newspapers}\label{tab:events}
\begin{tabular}{r|p{6.8cm}}
  18.08 & UK and France to sign Calais security deal. \\
  20-21.08 & Macedonian police teargas thousands of refugees crossing from Greece and declares state of emergency over surge in migrants \& refugees.  \\
  27-28.08 & 71 dead refugees found dead in truck in Austria. \\
  31.08 & Angela Merkel: Europe as a whole must help with refugees. \\
  1.09 & Hungary closes main Budapest station to refugees. \\
  2.09 & Alan Kurdi drowned off the shores of Turkey. \\
  4-6.09 & Migrants are allowed to cross the Austro-Hungarian border; Refugees welcomed warmly in Germany. \\
  8.09 & Hungarian Journalist appears to kick and trip fleeing refugees. \\
  14.09 & Austria followed Germany's suit and instituted border controls;  Refugee boat sinking: dozens including children drown off Greek island. \\
  15.09 & Croatia started to experience the first major waves of refugees; Hungary announced it would start arresting people crossing the border illegally. \\
  16.09 & Refugee crisis escalates as people break through Hungarian border; Hungary had detained 519 people and pressed criminal charges against 46 for trespassing, leading to pursue alternative routes through Croatia from Serbia. \\
  17.09 & Croatia decided to close its border with Serbia. \\
  
\end{tabular}
\end{table}

\begin{figure}[b]
	\centering
		\includegraphics[width=0.95\columnwidth]{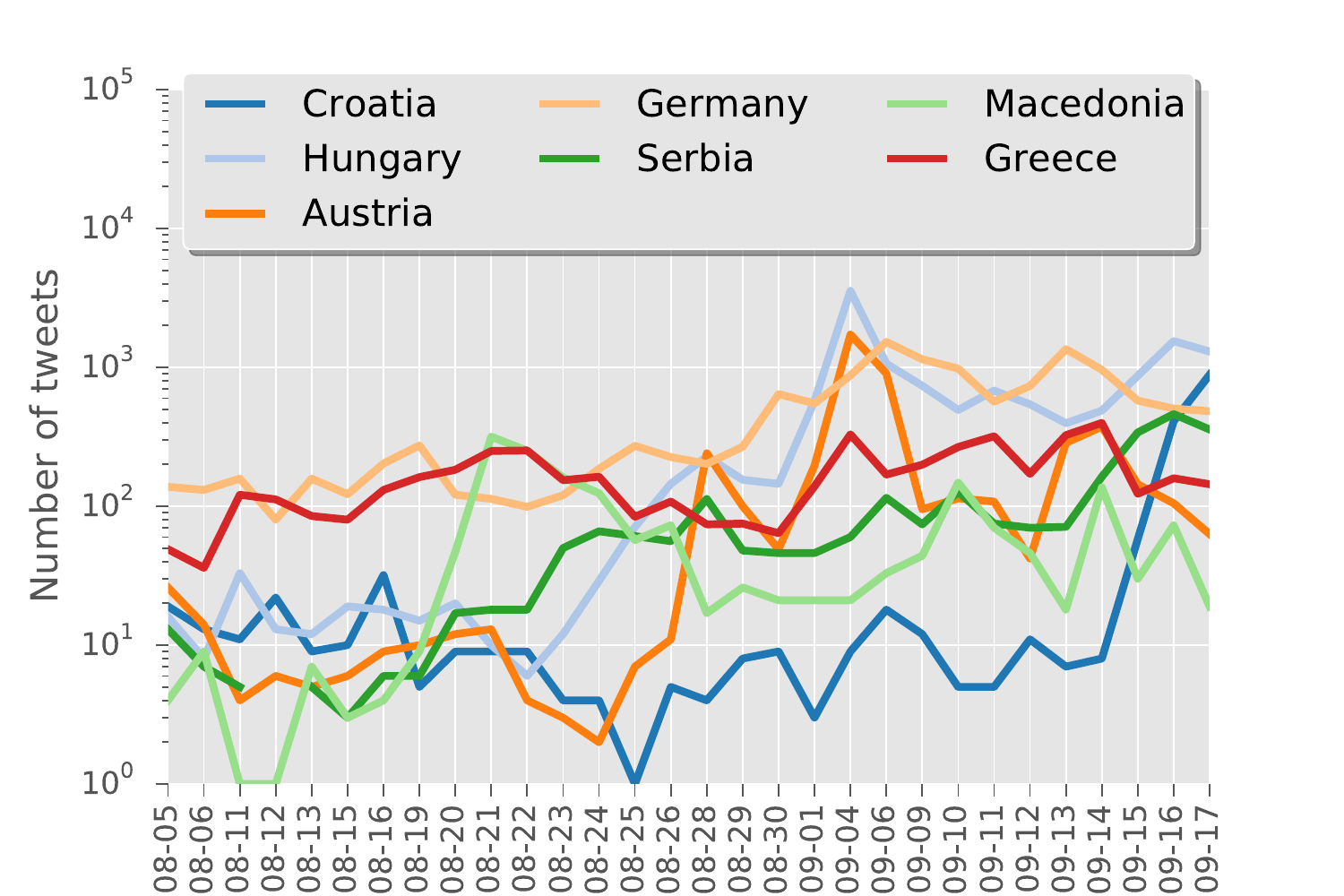}
	\caption{EU country mentions per day in log scale.}
	\label{fig:tweets-day-country-mentioned-EU}
\end{figure}

\begin{figure}[h]
	\centering
		\includegraphics[width=0.95\columnwidth]{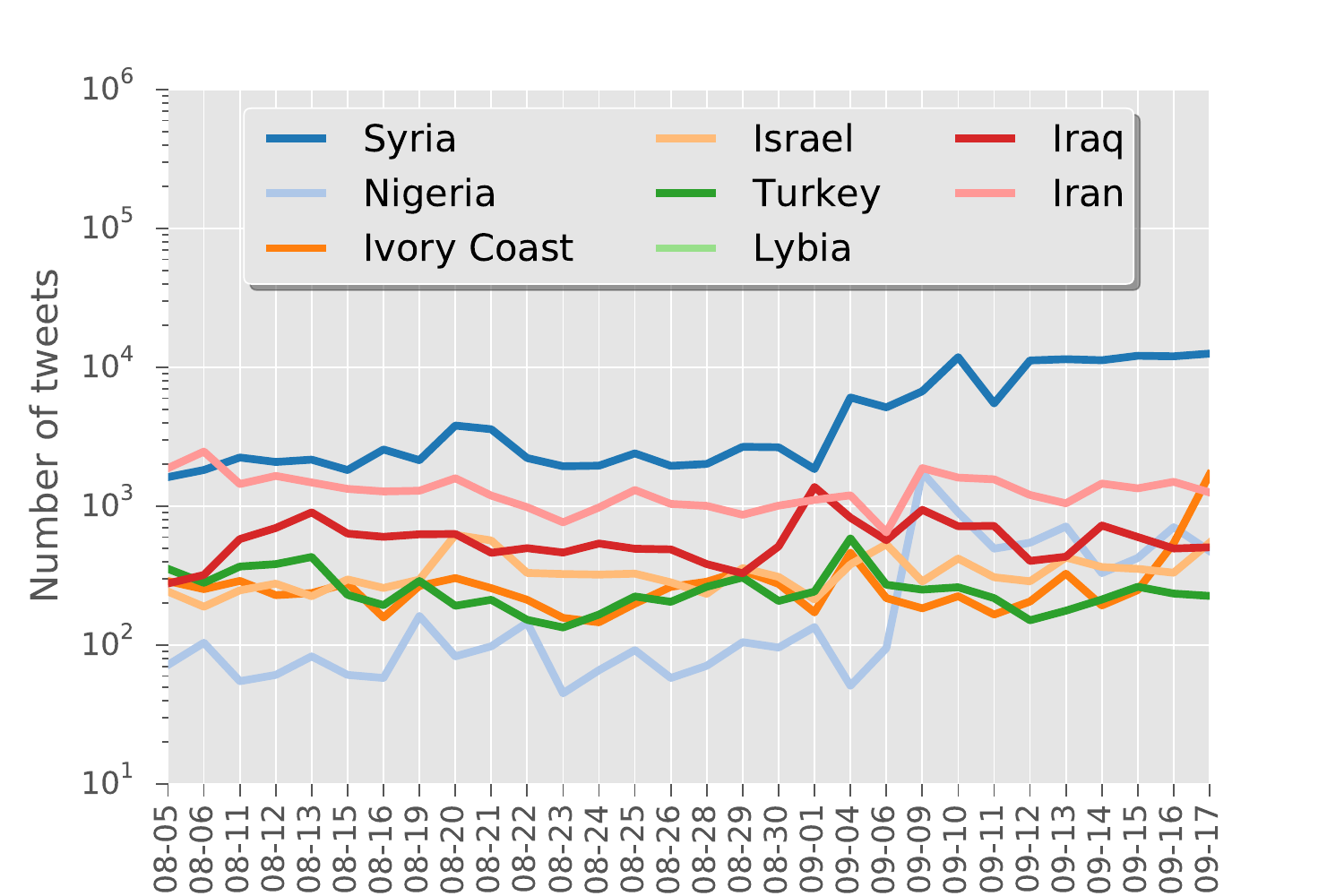}
	\caption{Non-EU country mentions per day in log scale.}
	\label{fig:tweets-day-mentioned-ORIENT}
\end{figure}


Next, we analyze location mention to a country related to the refugee  migration.  
Figure~\ref{fig:tweets-day-country-mentioned-EU}
reports the volumes of tweets mentioning the  EU countries most impacted by the refugees route, namely Austria, Germany, Croatia, Macedonia, Hungary, Serbia, Greece. We see that there is an interesting correspondence between the peaks of mentions and the events timeline. An evident peak for Germany, Austria and Hungary is the first week of September, probably related to the issues of borders being opened to refugees. We also notice a peak of mentions of Croatia corresponding to the close of borders with Serbia.  
Macedonia also sees and important increase of mentions around the 20 of August, probably in relation to the  Macedonian Police using gas towards refugees.

 
Similarly, Figure~\ref{fig:tweets-day-mentioned-ORIENT} focuses on the mention of relevant non European countries. The number of tweets mentioning Syria increases dramatically after the aforementioned facts of September 4, Turkey has a peak the day 4 of September due probably to the young kid found on the beach. We also observe how the mentions to other countries remains more or less stable along this period to witness the fact that they were not directly related to the events reported by the media in that period and that involved mainly the Syrian refugees. 

From a content standpoint we tracked how hashtags usage is closely related to the relevant events. 
We have counted the frequency of each hashtag in each day, then performed a two-pass normalization.
First we normalized the frequencies of hashtags on each day so as to avoid that days with lower recorded traffic are given less importance.
Then we normalized each hashtag over the observed period, so that the values are comparable among different hashtags.
We then measured the variance of the normalized frequencies, considering that hashtags with higher variance are those with a more unbalanced distribution among days. The hypothesis is that the unbalanced distribution is due to a close relation of the hashtag with a specific temporal event (usually one or two days).
Figure \ref{fig:hashtags} shows the resulting twenty highest-variance hashtags. 
The plot shows how this simple method allowed us to quickly spot hot topic in the observed stream of tweets and to correctly place them in time.

\begin{figure}[b]
	\centering
		\includegraphics[width=0.98\columnwidth]{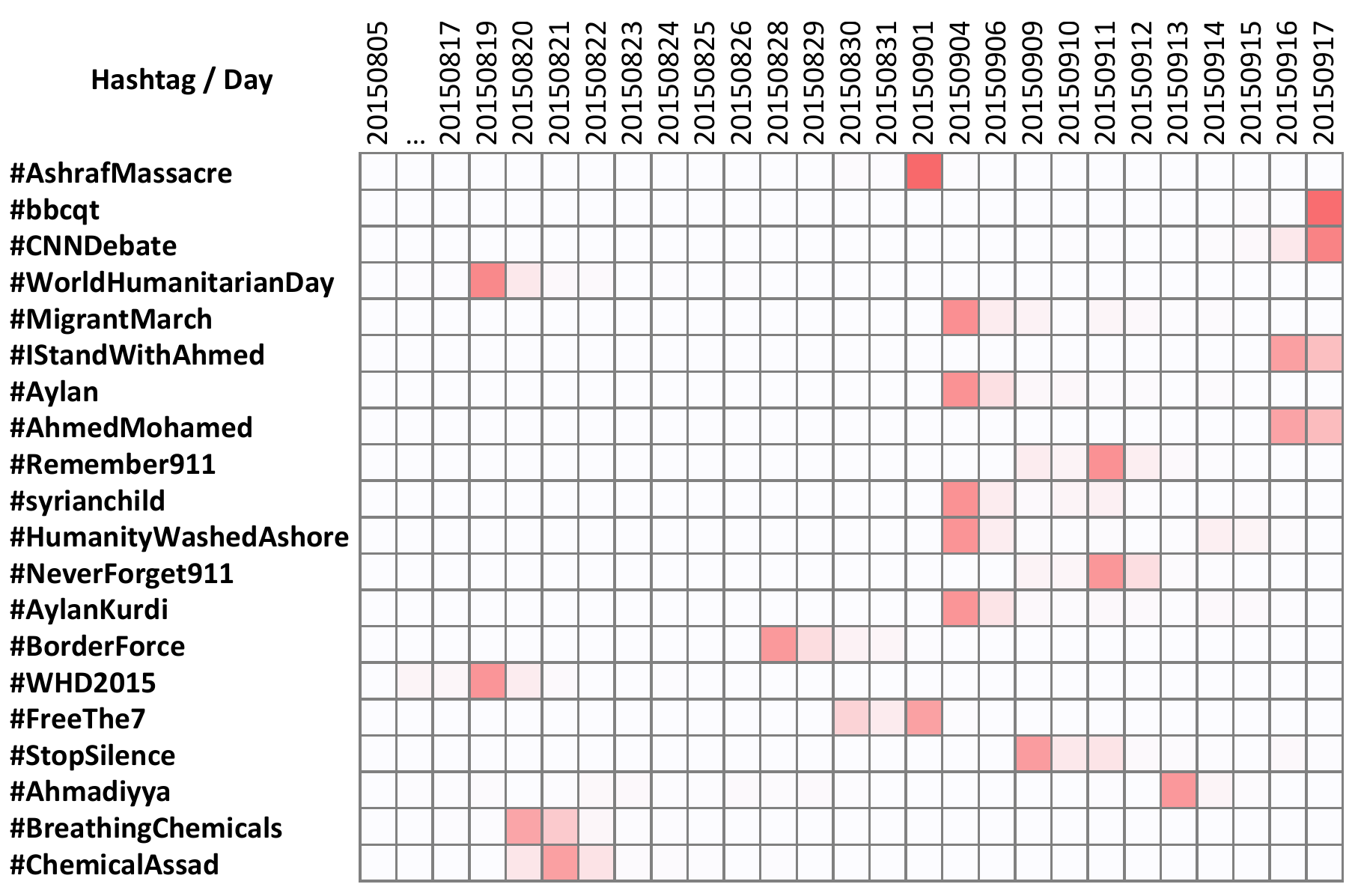}
	\caption{Highest-variance hashtags per day; intense red represents higher relative freq.}
	\label{fig:hashtags}
\end{figure}



%
%
%
%
%


\begin{figure*}[t!]
\begin{tabular}{@{}c@{}c@{}c@{}}
	\includegraphics[width=0.33\textwidth]{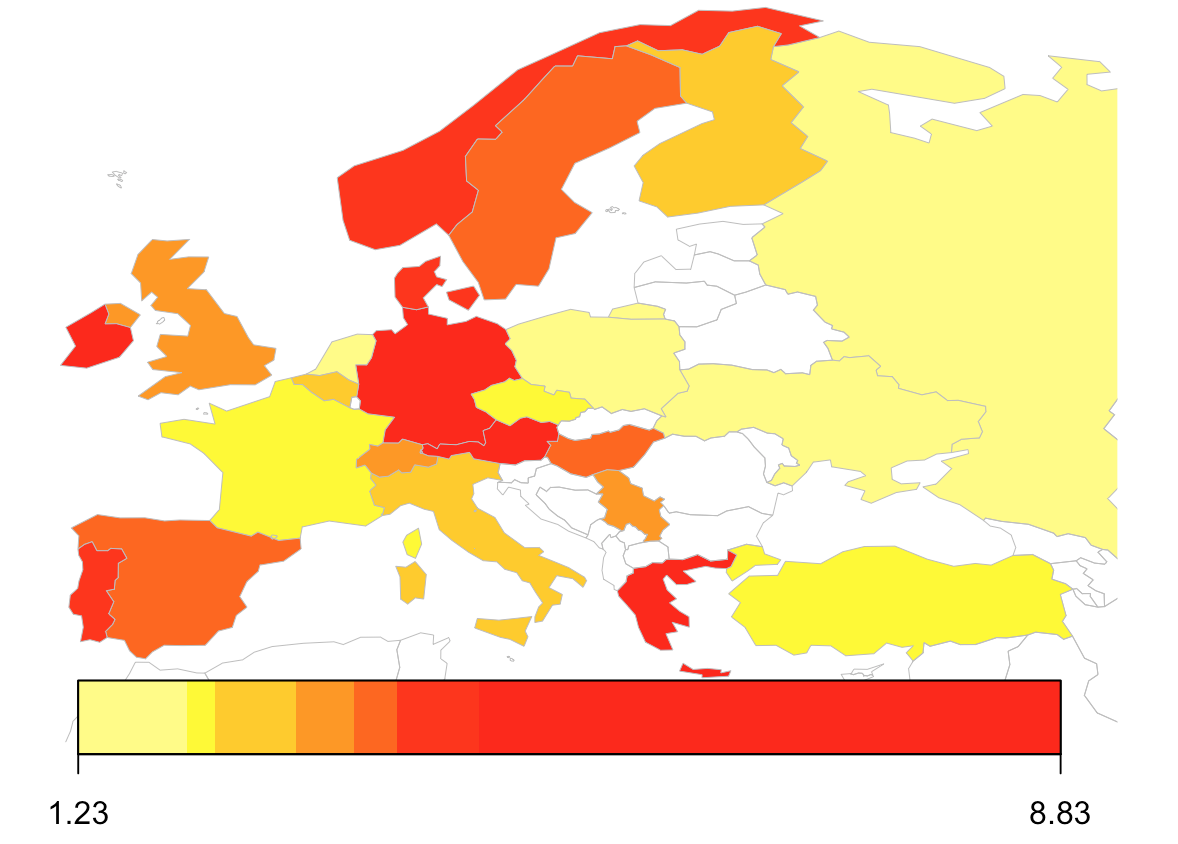} & 
	\includegraphics[width=0.33\textwidth]{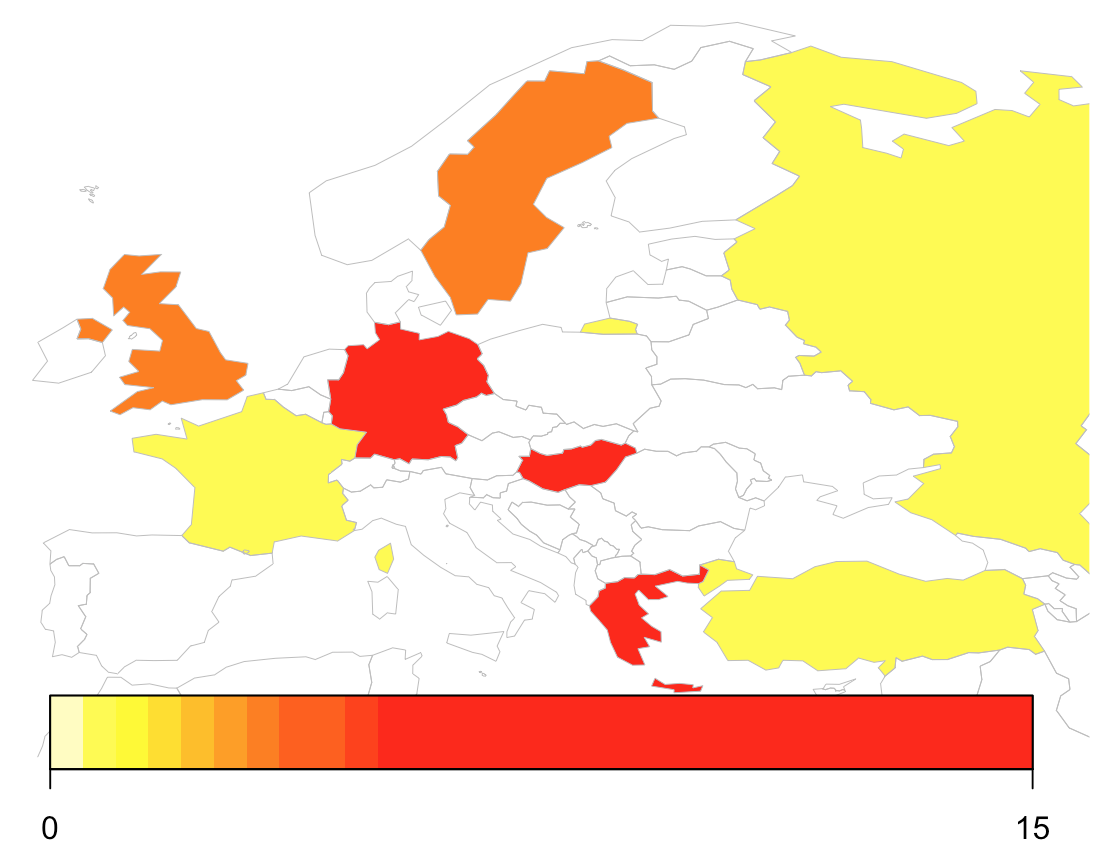} &
	\includegraphics[width=0.33\textwidth]{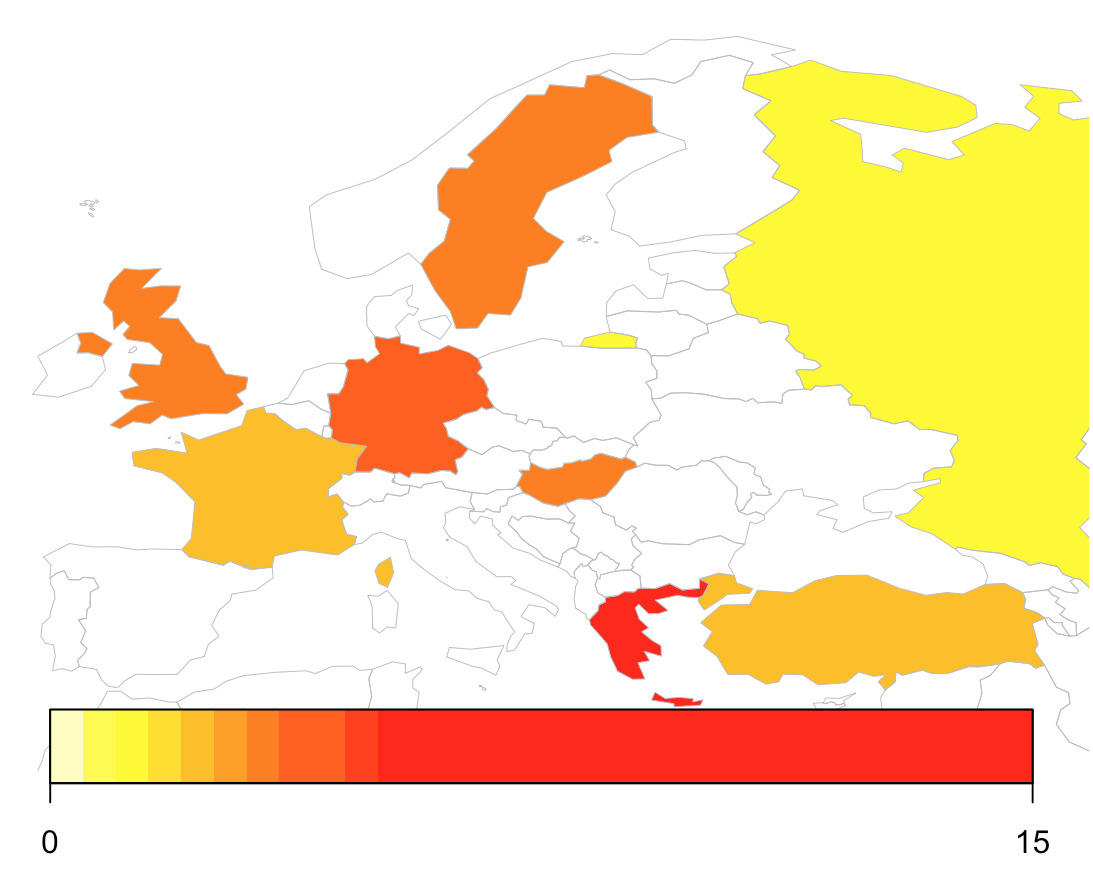} \\
	(a) Global perception & (b)  Internal perception & (c) External perception\\
\end{tabular}
	\caption{Index $\rho$ across European countries: red corresponds to a higher predominance of positive sentiment, yellow indicates lower $\rho$. (a) Refers to the whole dataset. (b) Is limited to users
	when mentioning locations in the their own country. (c) Is limited to users otherwise.}
	\label{fig:sentiment_us}
\end{figure*}

\subsection{Sentiment Analysis}
\label{subsec:paruserlco}


To answer the analytical question AQ2, we analyze the perception of the refugee crisis phenomenon by the European countries
by exploiting the sentiment and location dimensions of the Twitter users
in our dataset.
To simplify the notation in the following we refer to \userloc simply by \users.  

Let us define $\rho$ the ratio between the number of polarized users pro \textit{refugees} and the number of users against \textit{refugees}:
\begin{equation} 
\nonumber
\rho = \frac{ |\users_{c_+}| }{ |\users_{c_-}|  }
\end{equation} 
The index $\rho$ gives a compact indication of the sentiment of a group of users. 

We first analyze the sentiment across the various countries, and we then differentiate
between discussion about internal and external locations.



%
%
%

\subsubsection{Sentiment by country}



Figure~\ref{fig:sentiment_us}(a) shows the value of $\rho$ for users belonging
to the different European countries.
We observe that Eastern countries in general are less positive than Western countries. In particular, Russia and Turkey have a low sentiment index probably because they are highly affected by the flow of arrivals. 
On the contrary, countries like Germany and Austria are more positive and this can be confirmed by the news  reporting their decision of opening borders to migrants. 
Among western countries, France, UK, Italy and Netherlands have a low $\rho$ index. In Italy the large amount of refugees arrived mainly through the sea directly from Lybia or Tunisia and the tone of the discussion is  often characterized by negative notes.  In France the sentiment confirms all worries about, probably,  the situation of the Calais jungle.
The situation in Greece appears very different. The sentiment is positive even though this country remains by far the largest single entry point for new sea arrivals in the Mediterranean, followed by Italy. Greece  
captured the attention of humanitarian organizations.
Countries like Ireland, Norway or Portugal are less interested by the phenomenon and therefore their perception might result more positive. Even for Spain $\rho$ is not particularly low since the problem of refugees coming from Western Mediterranean was limited in number of people compared to central and eastern countries. 

\subsubsection{Internal and external country perception of the refugees crisis}

In the following we study the perceived sentiment in relation to the user country. 
We denote as {\em internal perception} the sentiment of a user when mentioning his/her own country (or a city in his/her country). {\em External perception} refers to polarized tweets with no internal references. 

Figure~\ref{fig:sentiment_us}(b) shows the sentiment ratio $\rho$ by country considering the internal perception, thus tweets mentioning the country itself. The $\rho$ computation is this case refers to the users of a country who mentioned in their tweets the country itself (or indirectly a city in the country).
We report only the countries for which we have a minimum amount of data. 
We can see that Russia, France and Turkey have a really low $\rho$ index. We conjecture that the sentiment of a person, when the problem involves directly his/her own country, could be more negative since we are generally more critical when issues are closer to ourselves. 

The external perception ratio is depicted in Figure~\ref{fig:sentiment_us}(b). Comparing the two maps, we see that internal and external perception is stable for UK and Sweden. Other countries have a much lower internal sentiment $\rho$ than external, and this is the case of France, Russia and Turkey. All these countries were indeed facing many critical problems due to the arrival of refugees to their borders. The case of \textit{Calais} is one of the most significant examples which could explain the case of the low ratio in France. 
Germany, Hungary and Greece, on the contrary, have a better internal perception which might be due to the decision of Germany to open borders to allow many people to transit from Hungary to Germany, releasing the extremely difficult situation at the national borders.

%

%
%
%
%
%

\begin{figure*}[t]
	\centering
		\includegraphics[width=0.93\textwidth]{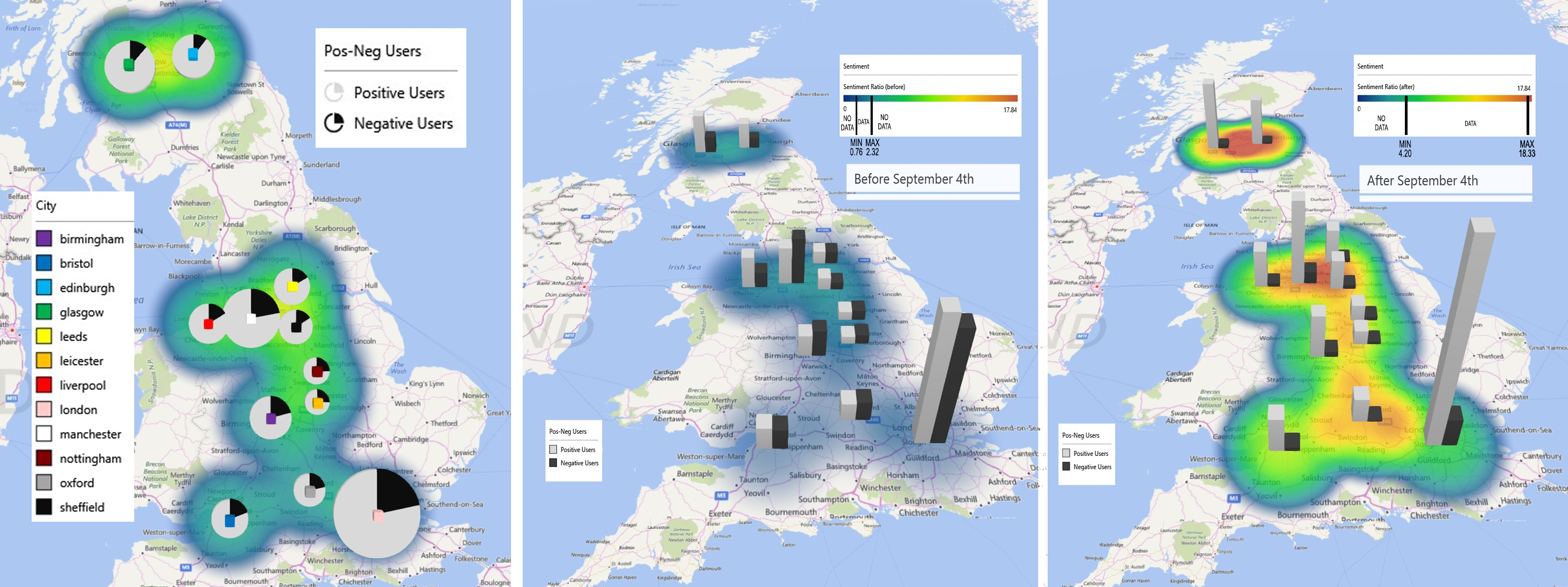}
	\caption{Positive and negative users for different cities in UK in all period (left) before (center) and after (right) September 4. In the infographic the pies/bars show the number of polarized positive and negative users by city and the heat map in background indicates the value of $\rho$ for the cities considered in the legend. For some cities the tweets are not sufficient to compute polarization, therefore when the heat degrades to 0 it indicates no data.}
	\label{fig:uk_bef_aft}
\end{figure*}

\subsubsection{Sentiment Analysis: the UK case}
\label{sec:uk}


In this section we focus on  the sentiment analysis of UK citizens,
as UK is the most represented country in our dataset 
and therefore a more detailed sentiment analyses can be done.
Indeed, for UK we detail the results at the granularity of the city level. 

\inote{aiuto non capisco userloc non erano gli utenti georefereniati?? sono tutti polarizzati }

Figure~\ref{fig:uk_bef_aft} (left) shows the number of polarized users \userloc in the most represented cities of the country, with at least 100 polarized users in the dataset. We can see from the heatmap that there is a gradient of polarization from south to north in the sentiment. This could be due to the fact that the cities in the south, and London in particular, were more involved in the welcoming process of refugees and this might have generated more discontent even though the general perception is quite positive. 
From the time series of $\rho$ for UK users we see an increase in the general sentiment ratio of the country after September 4. 
We find news\footnote{Sept, 04: \url{http://www.bbc.com/news/uk-34148913} -- Sept, 16: \url{http://www.bbc.com/news/uk-34268604}} regarding that period from BBC and we think that the increase in the sentiment polarization could be be due mainly to the decision of the Prime Minister Cameron of acting with ``head and heart'' to help refugees. He allocated substantial amounts of money to humanitarian aid becoming, at that time, the second largest bilateral donor of aid to the Syrian conflict (after the US).
Figure~\ref{fig:uk_bef_aft} (middle) and (right), shows the comparison of the opinion in UK before and after September 4, respectively. We highlight again a gradient of polarization from north to south in both cases even though the sentiment ratio $\rho$ before and after that day is completely overturned. After September 4 the spreading of positive news in UK increases the sentiment and the volume of relevant tweets in all the country and probably government position reflects the sentiment of a vast majority of users which show support to refugees in their digital statements.

\enlargethispage*{1cm}

\subsection{Mentioned Location Analysis}
The last analysis we conducted aims at answering AQ3 by studying the sentiment of the tweets when mentioning specific countries.
We show how events impact differently the volume of tweets with positive or negative
sentiment. Furthermore, we relate the sentiment changes to events.



\begin{figure*}[t]
\centering     
\subfloat[Tweet sentiment for Hungary]{\label{fig:mentionHUsenti}\includegraphics[width=0.32\textwidth]{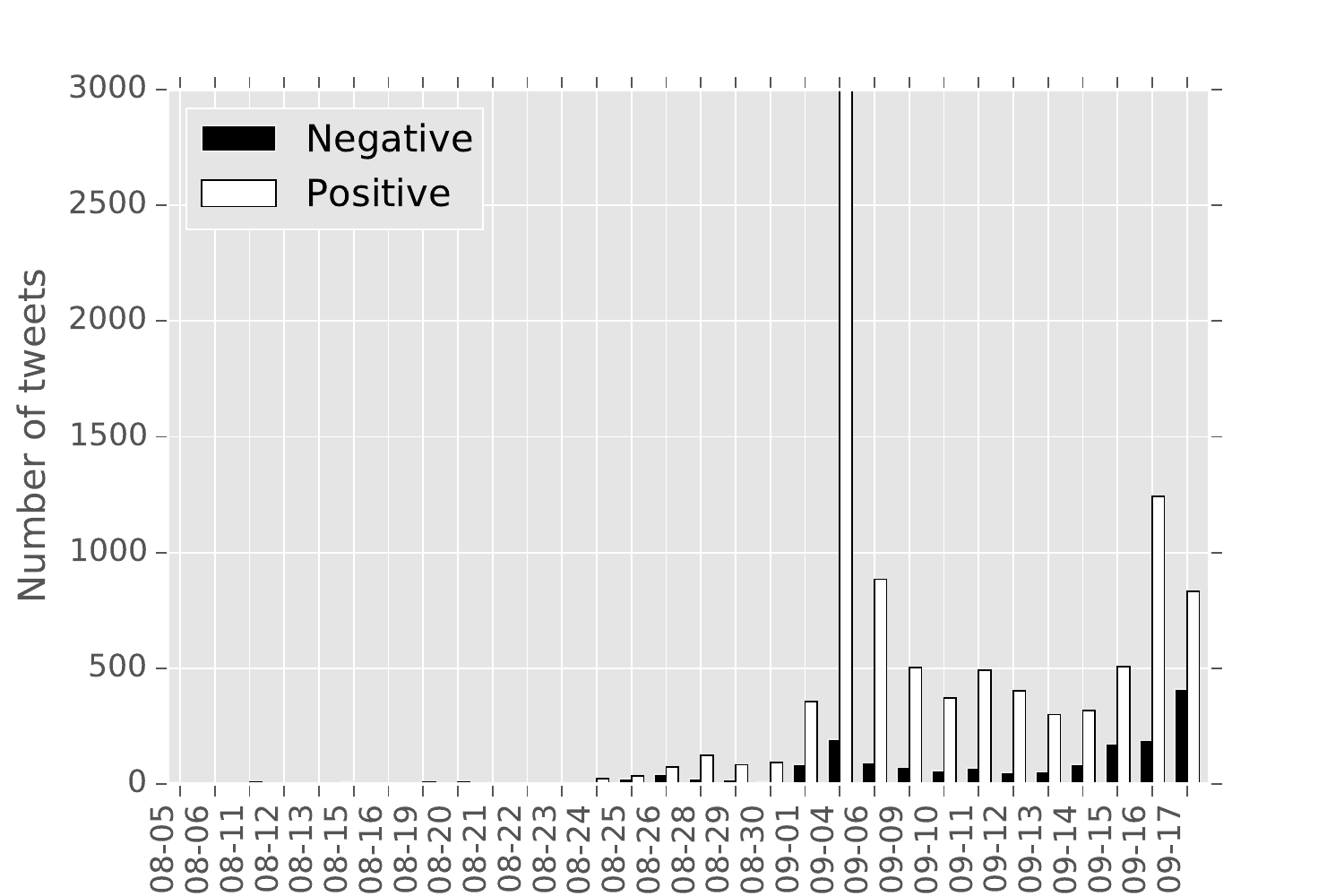}}
\subfloat[Tweet sentiment for Austria]{\label{fig:mentionATsenti}\includegraphics[width=0.32\textwidth]{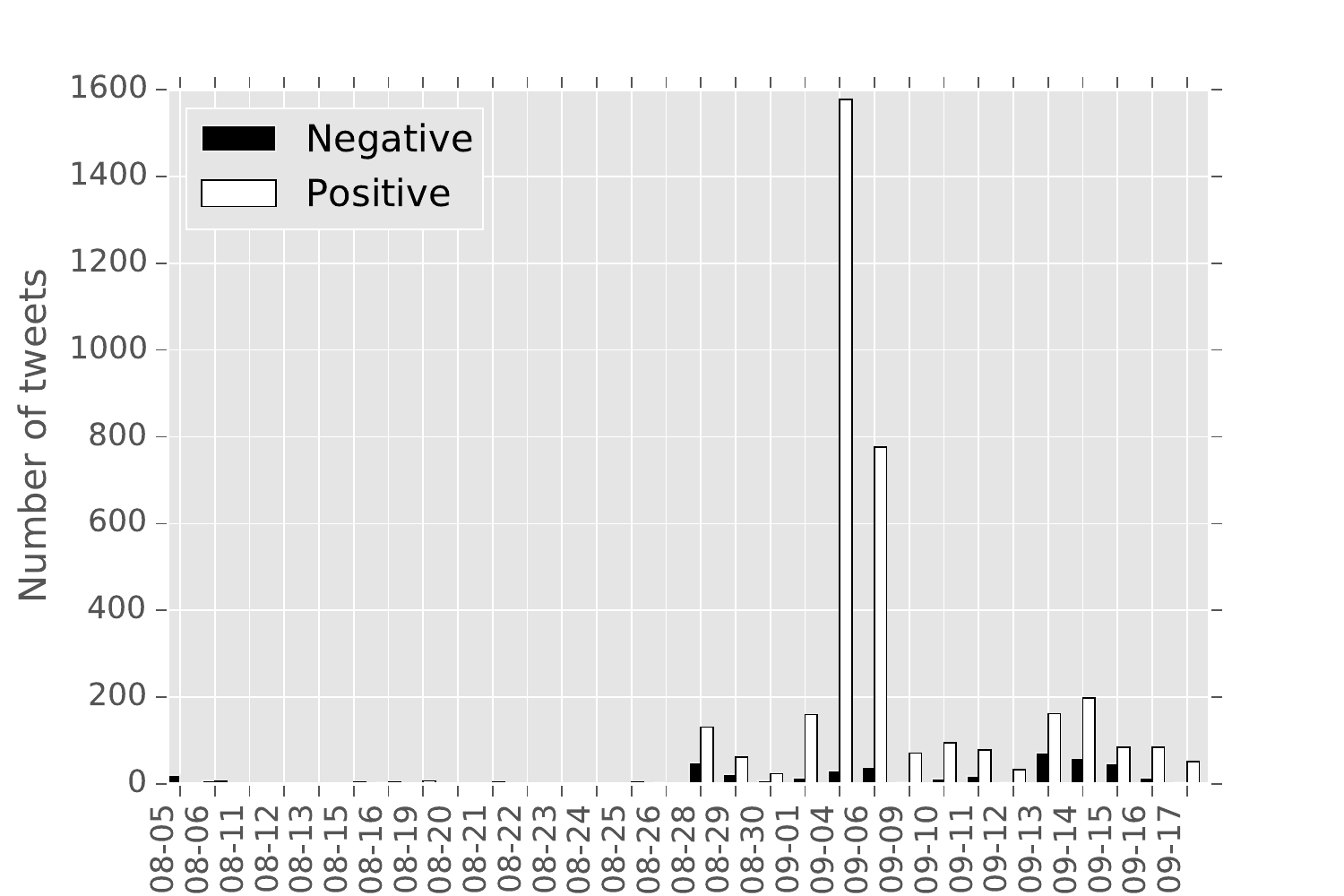}}
\subfloat[Tweet sentiment for Croatia]{\label{fig:mentionCRsenti}\includegraphics[width=0.32\textwidth]{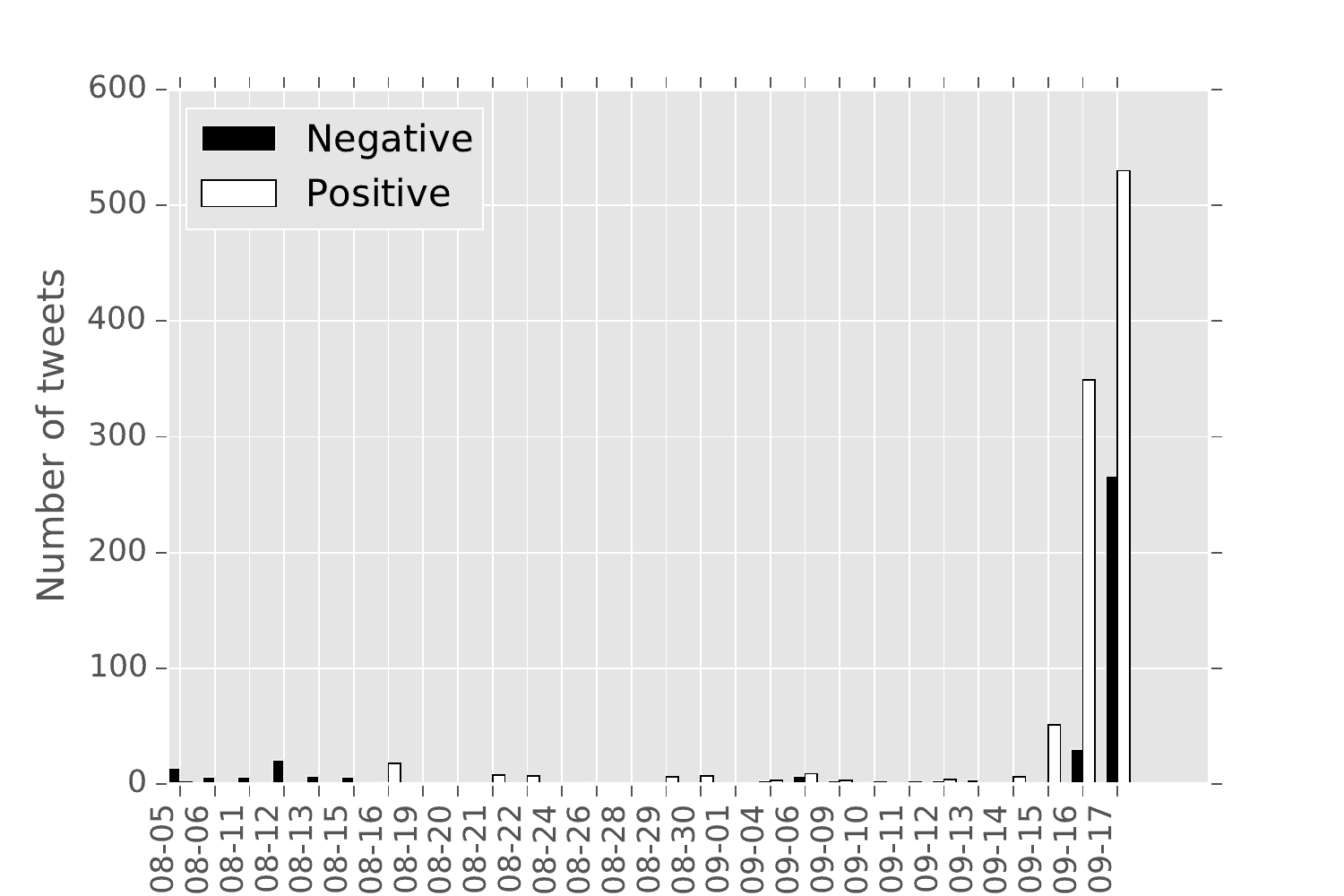}} \\
\subfloat[Tweet sentiment for UK]{\label{fig:mentionUKsenti}\includegraphics[width=0.32\textwidth]{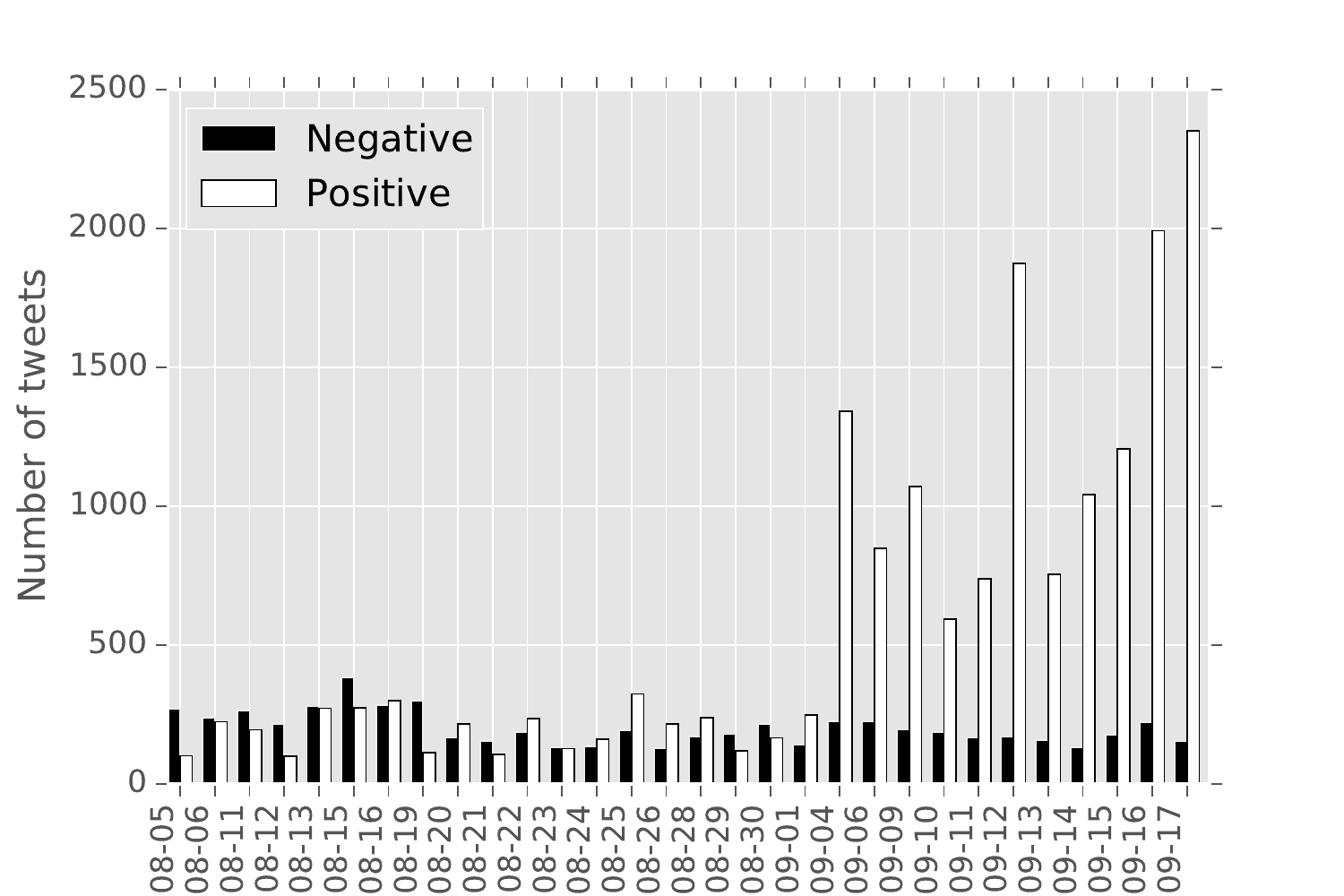}}
\subfloat[Tweet sentiment for France]{\label{fig:mentionFRsenti}\includegraphics[width=0.32\textwidth]{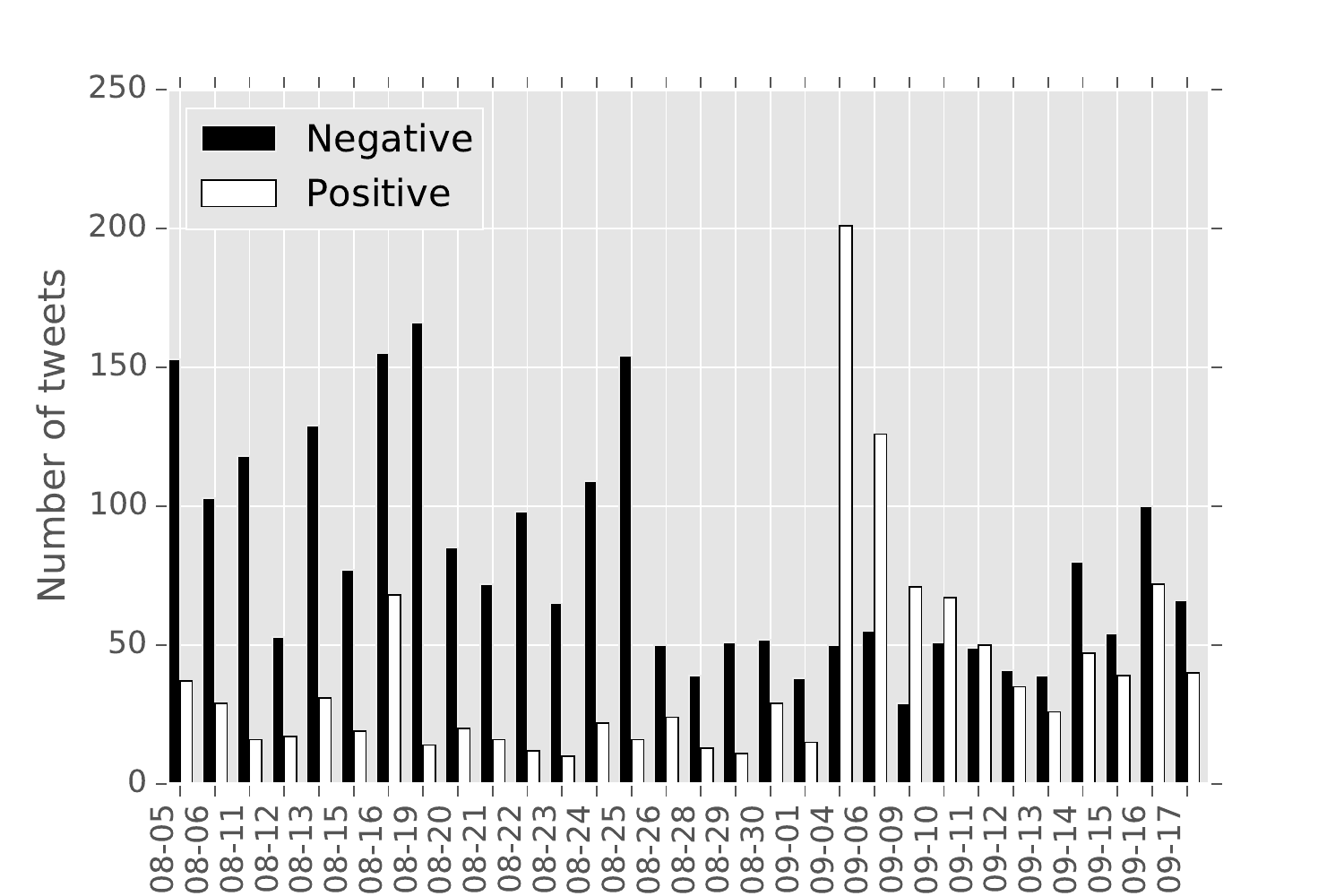}}
\subfloat[Tweet sentiment for Germany]{\label{fig:mentionDEsenti}\includegraphics[width=0.32\textwidth]{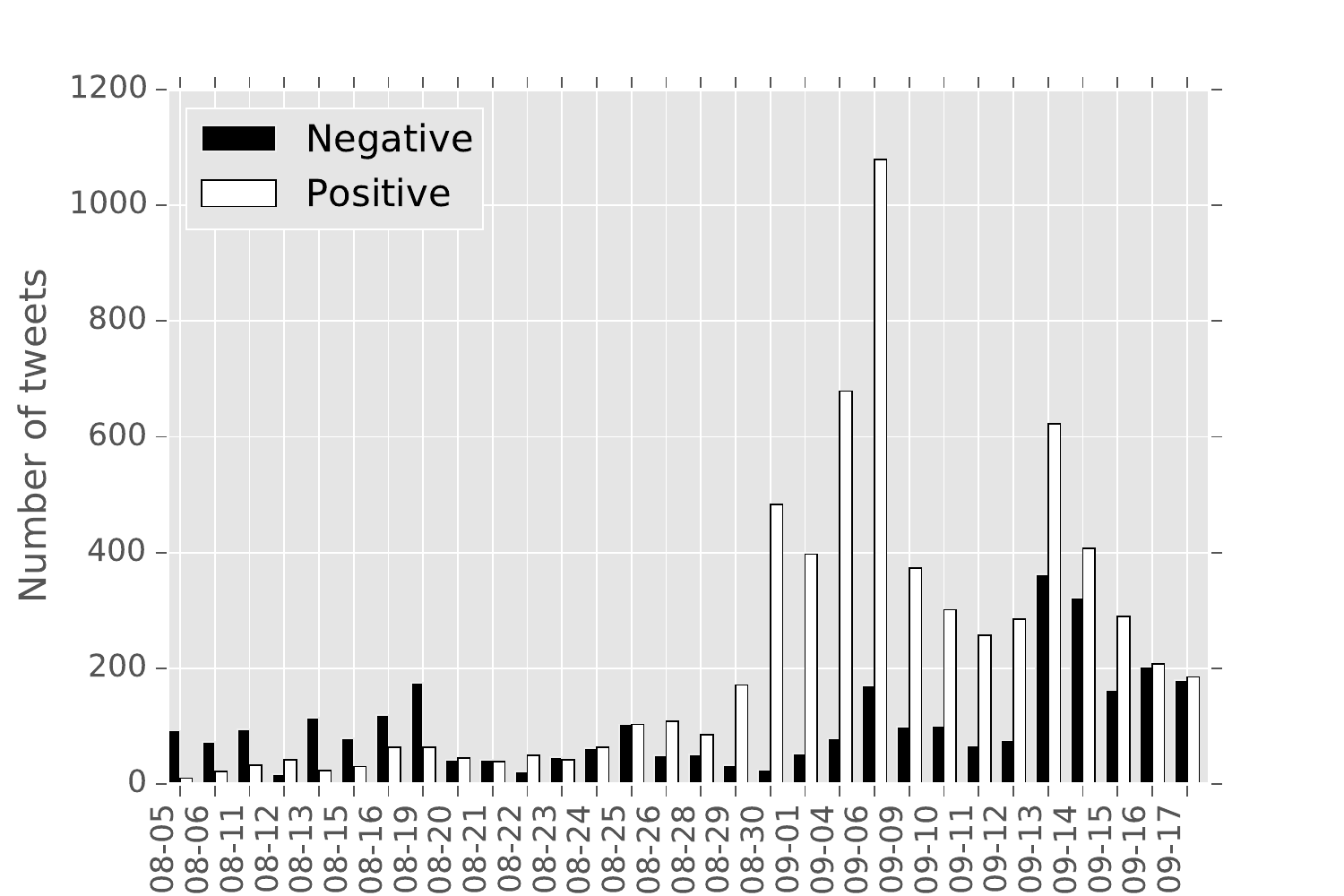}} 
\caption{Tweet sentiment for country mentions per day.}
\label{fig:featureAnalysis}
\end{figure*}

Figure~\ref{fig:featureAnalysis}~(a-c) shows the sentiment of tweets when mentioning three of the countries most impacted
by the refugees routes: Hungary, Austria and Croatia. We highlight an overall low number of mentions to these countries until the beginning of September. In the case of Hungary  and Austria there is a sudden increase at the beginning of September in the overall number of mentions, predominantly for $c_+$ with a relative increase in $c_-$. This is mostly due to the overall positive sentiment towards the events from the previous days (the death of Alan Kurdi, the young boy found he drowned), but also due to the good news of migrants being allowed to cross the Austro-Hungarian border. The negative sentiment starts to appear, and continues to grow, until the middle of September when the negative class tends to increase more than the positive one. Here we have tweets expressing negative feelings towards border controls on behalf of Austria (13-15 Sept) and Hungary arresting refugees crossing the border illegally (15-17 Sept). Croatia comes into play towards the end of our observation period when on the September 16th becomes a valid alternative to Hungary which closed its borders with Serbia. A similar analysis has been done for Greece, Macedonia and Serbia, all countries facing critical moments, but due to lack of space we are omitting here.

%
%
In Figure \ref{fig:featureAnalysis} (d-f) we look at the sentiment in relation to the mentions of UK, France and Germany. Both UK and Germany are rather balanced between positive and negative tweets. Germany presents exceptions on certain days when a positive feeling arises in support to the sad incidents related to refugees. We notice that the official media news at the end of August reported that Germany was welcoming refugees, while UK started showing a positive sentiment after the dramatic facts of Alan Kurdi and the announcement of taking in 20,000 refugees by 2020. France seems to have more negative feelings, probably due to the difficult situation in Calais and news about victims trying to across to UK while a positive peak appears in correspondence to the young victim found.



\section{Conclusions and Future Work}
\label{sec:conclusions}
\enlargethispage*{1cm}
\vspace{-0.25cm}

We proposed an adaptive and scalable multidimensional framework to analyze the spatial, temporal and sentiment aspects of a polarized topic discussed in an online social network. As a case study we used a Twitter dataset related to the Mediterrean refugee crisis. Besides enriching tweets with spatial and temporal information, a contribution of this paper is the sentiment enrichment able to identify the polarity of users and tweets.
The combination of the sentiment aspects with the temporal and spatial dimension is an added value that allows us to infer interesting  insights. 
Our analysis revealed that European users are sensitive to major events and mostly express positive sentiments for the refugees, but in some cases this attitude suddenly changes when countries are exposed more closely to the migration flow.
As a future work we intend to adapt the framework to a real-time streaming scenario and to add more dimensions such as the type of user and the network relationships in the Twitter user graph.


{\bf Acknowledgments. }
{\small This work was partially supported by the EC H2020 Program INFRAIA-1-2014-2015 {\em SoBigData: Social Mining \& Big Data Ecosystem} (654024).}



\newcommand{\BIBdecl}{\setlength{\itemsep}{0.25 em}}
\bibliographystyle{IEEEtran}
%
%
%


\vspace{-0.25cm}
\bibliography{biblio}

\end{document}